\documentclass[aps,prd,floatfix,amsfonts,amssymb,amsmath,a4paper,twocolumn,notitlepage,nofootinbib,groupedaddress,superscriptaddress]{revtex4-1}
\usepackage{amsmath,amssymb}
\usepackage{graphicx}
\usepackage{slashed}
\usepackage[T1]{fontenc}
\usepackage{color}
\usepackage{hyperref}
\usepackage{times}
\usepackage{overpic}
\usepackage{rotating}

\newcommand{\e}{\operatorname{e}}
\newcommand{\stext}[1]{\raisebox{-1.5pt}{{\scriptsize #1}}} % for subscripts
\definecolor{tumblau}{cmyk}{1,0.43,0,0}
\definecolor{tumrot}{RGB}{202,033,063}
\definecolor{tumgruen}{cmyk}{0.35,0,1,0.2}

\begin{document}
\title{From asymmetric nuclear matter to neutron stars: a functional renormalization group study}
\author{Matthias Drews}
	\affiliation{Physik Department, Technische Universit\"{a}t M\"{u}nchen, D-85747 Garching, Germany}
	\affiliation{ECT*, Villa Tambosi, I-38123 Villazzano (Trento), Italy}
\author{Wolfram Weise}
	\affiliation{Physik Department, Technische Universit\"{a}t M\"{u}nchen, D-85747 Garching, Germany}
	\affiliation{ECT*, Villa Tambosi, I-38123 Villazzano (Trento), Italy}
\date{\today}

\begin{abstract}
A previous study of nuclear matter in a chiral nucleon-meson model is extended to isospin-asymmetric matter. Fluctuations beyond mean-field approximation are treated in the framework of the functional renormalization group. The nuclear liquid-gas phase transition is investigated in detail as a function of the proton fraction in asymmetric matter. The equations of state at zero temperature of both symmetric nuclear matter and pure neutron matter are found to be in good agreement with realistic many-body computations. We also study the density dependence of the pion mass in the medium. The question of chiral symmetry restoration in neutron matter is addressed; we find a stabilization of the phase with spontaneously broken chiral symmetry once fluctuations are included. Finally, neutron star matter including beta equilibrium is discussed. The model satisfies the constraints imposed by the existence of two-solar-mass neutron stars.
\end{abstract}

\maketitle

\section{Introduction}
Recent high-quality observational data of neutron stars set new stringent constraints for the equation of state (EoS) of cold and dense matter, otherwise inaccessible by experiment. The masses of two heavy pulsars were determined with high precision: J1614-2230 with \mbox{$M=(1.97\pm0.04)\,M_\odot$} \cite{demorest2010two-solar-mass}, and J0348+0432 with \mbox{$M=(2.01\pm0.04)\,M_\odot$} \cite{antoniadis2013massive}. Only a sufficiently stiff EoS can support such neutron stars against gravitational collapse. While neutron star radii are much less accurately known, the combination of available data makes these objects nonetheless an indispensable tool to constrain possible equations of state \cite{truemper2011observations,hebeler2013equation,hell2014dense}.

Theoretical investigations of neutron-rich matter have quite well converged in recent years. Different approaches, such as chiral effective field theory (ChEFT \cite{fritsch2005chiral,fiorilla2012chiral,holt2013nuclear}), chiral Fermi liquid theory \cite{holt2013chiral}, as well as Quantum Monte Carlo (QMC) methods, using either phenomenological interactions \cite{gandolfi2014equation,gandolfi2012maximum} or ChEFT potentials \cite{gezerlis2013quantum,roggero2014quantum,holt2014auxiliary-field}, agree well in their common overlap region of applicability at lower densities.

At higher densities, effects of three-body forces and higher-order pion-exchange processes become increasingly important \cite{hebeler2010chiral,tews2013neutron}, and it is crucial that any realistic model takes into account fluctuations and correlations generated by those mechanisms. A powerful method to study the effects of fluctuations in a consistent and fully non-perturbative way is the functional renormalization group (FRG \cite{berges2002non-perturbative,pawlowski2007aspects,gies2012introduction,delamotte2012introduction,braun2012fermion}).

In recent studies \cite{drews2013thermodynamic,drews2014dense,drews2014functional} we have applied FRG methods to a chiral nucleon-meson model \cite{berges2000chiral,berges2003quark} for symmetric nuclear matter and pure neutron matter.
The present paper extends these calculations to asymmetric nuclear matter, systematically varying the relative proportion of neutron and proton densities. In Section~\ref{sec:model} we review the setup of the model and explain how the parameters are adjusted in a mean-field calculation. Section~\ref{fluctuations} demonstrates how to include fluctuations using the framework of the functional renormalization group. We then study in Section~\ref{sec:asymmetric} the nuclear liquid-gas phase transition and the equation of state for varying proton fractions. Section~\ref{sec:pionmass} deals with  the in-medium pion mass as a useful test observable for pionic fluctuations. In Section~\ref{sec:chiral_restoration} we discuss the issue of chiral symmetry restoration with emphasis on neutron matter. Finally, in Section~\ref{sec:neutronstars} neutron star matter under the condition of beta equilibrium is examined and discussed in view of the new observational constraints.

\section{Extended chiral nucleon-meson model}\label{sec:model}
In the hadronic phase of QCD with spontaneously broken chiral symmetry, the active degrees of freedom are baryons and mesons (predominantly nucleons and pions). Given that the equation of state of cold and dense baryonic matter must be sufficiently stiff to support two-solar-mass neutron stars, this constraint implies limitations on the baryon densities that can be reached in the center of the star. Several studies utilizing the observational constraints typically find central densities not exceeding about five times nuclear saturation density, \mbox{$n_0=0.16\text{\,fm}^{-3}$}~\cite{hebeler2013equation,gandolfi2014equation,hell2014dense}. Under such conditions, exotic compositions with substantial portions of quark matter or kaon condensates are unlikely. Also, the appearance of hyperons would make the EoS too soft unless strongly repulsive correlations are introduced to sustain sufficiently large pressure gradients at high densities \cite{akmal1998equation,weissenborn2012hyperons,lonardoni2014hyperon}. In this work we will not consider such effects and restrict ourselves to ``conventional'' nucleon and meson degrees of freedom.

The present approach uses a chiral nucleon-meson (ChNM) model~\cite{berges2000chiral,berges2003quark,floerchinger2012chemical} which has been demonstrated to be suitable for studying the thermodynamics of baryonic matter. It is based on a \mbox{$\operatorname{SU}(2)_L\times\operatorname{SU}(2)_R$} linear sigma model. Protons and neutrons are combined in an isospin doublet field~\mbox{$\psi=(\psi_p,\psi_n)^{\text T}$} and coupled to isoscalar-scalar ($\sigma$) and pion ($\boldsymbol\pi$) fields transforming as a four-component field $(\sigma,\boldsymbol\pi)^{\text T}$ under the chiral group. Long and intermediate range interactions of the nucleons are generated by $\boldsymbol\pi$ and $\sigma$ exchange mechanisms. The nucleon-nucleon forces at short distance are conveniently described in terms of four-fermion isoscalar- and isovector-vector current interactions, \mbox{$(\bar\psi\gamma_\mu\psi)\,(\bar\psi\gamma^\mu\psi)$} and \mbox{$(\bar\psi\gamma_\mu\boldsymbol\tau\psi)\cdot(\bar\psi\gamma^\mu\boldsymbol\tau\psi)$}, respectively, where $\boldsymbol\tau$ are the isospin Pauli matrices. A Hubbard--Stratonovich transformation bosonizes these interactions, introducing effective vector-isoscalar and vector-isovector fields, $\omega_\mu$ and $\boldsymbol\rho_\mu$, respectively. In the present study, while fluctuations of the $\boldsymbol\pi$ and $\sigma$ fields will be included non-perturbatively, $\omega_\mu$ and $\boldsymbol\rho_\mu$ are treated as mean fields. These vector bosons conveniently parametrize unresolved short-distance physics. They are not to be identified with the physical omega and rho mesons. The Lagrangian (in Minkowski space-time) of this extended chiral nucleon-meson model reads:
\begin{align}\label{eq:Lagrangian}
	\begin{aligned}
		\mathcal L_{\stext{ChNM}}&=\bar\psi i\gamma_\mu\partial^\mu\psi+\frac 12\partial_\mu\sigma\,\partial^\mu\sigma+\frac 12\partial_\mu\boldsymbol\pi\cdot\partial^\mu\boldsymbol\pi \\
		&\quad-\bar\psi\Big[g(\sigma+i\gamma_5\,\boldsymbol\tau\cdot\boldsymbol\pi)+\gamma_\mu(g_\omega\, \omega^\mu+g_\rho\boldsymbol\tau\cdot\boldsymbol\rho^\mu)\Big]\psi \\
		&\quad-\frac 14 F^{(\omega)}_{\mu\nu}F^{(\omega)\mu\nu} - \frac 14 \boldsymbol F^{(\rho)}_{\mu\nu}\cdot\boldsymbol F^{(\rho)\mu\nu} \\
		&\quad+\frac 12m_v^2\big(\omega_\mu\,\omega^\mu+\,\boldsymbol\rho_\mu\cdot\boldsymbol\rho^\mu\big)- {\cal U}(\sigma,\boldsymbol\pi).
	\end{aligned}
\end{align}
The field strength tensors of the vector bosons $\omega_\mu$ and $\boldsymbol\rho_\mu$ are generally given as \mbox{$F_{\mu\nu}^{(\omega)}=\partial_\mu\omega_\nu-\partial_\nu\omega_\mu$} and \mbox{$\boldsymbol F_{\mu\nu}^{(\rho)}=\partial_\mu\boldsymbol\rho_\nu-\partial_\nu\boldsymbol\rho_\mu-g_\rho\,\boldsymbol\rho_\mu\times\boldsymbol\rho_\nu$}, respectively. Treated as background fields, only $\omega_0$ and $\rho_0^3$ are non-vanishing and the non-abelian part of $\boldsymbol F_{\mu\nu}^{(\rho)}$ does not contribute in practice. The mass parameter $m_v$ associated with the vector fields encodes the characteristic scale of unresolved short-distance dynamics. Phenomenological boson-exchange models often use \mbox{$m_v\simeq0.8\text{\,GeV}$}. In the present approach only the effective coupling strengths of dimension (length)${}^2$, 
\begin{align}
G_\omega={g_\omega^2\over m_v^2}~,~~~ G_\rho={g_\rho^2\over m_v^2}~,\nonumber
\end{align}
are relevant. If the vector fields are integrated out, these couplings correspond to the respective local four-fermion NN interactions.

The microscopic potential, \mbox{$\mathcal U(\sigma,\boldsymbol\pi)$}, can be decomposed into a chirally invariant piece, \mbox{$\mathcal U_0(\chi)$}, which depends only on the chirally invariant square,
\begin{align}
\chi=\frac 12(\sigma^2+\boldsymbol\pi^2)\,\,,\nonumber	
\end{align} 
and an explicit symmetry breaking term:
\begin{align}
	\mathcal U(\sigma,\boldsymbol\pi)=\mathcal U_0(\chi)-m_\pi^2f_\pi(\sigma-f_\pi)\,,
\end{align}
where $m_\pi$ and $f_\pi$ are the physical mass and decay constant of the pion (in practice we use \mbox{$m_\pi=135\text{ MeV}$} and \mbox{$f_\pi=93\text{ MeV}$} as in Ref.~\cite{floerchinger2012chemical}).

Finite temperatures and chemical potentials are treated within the Matsubara formalism. The time-component is Wick-rotated, $t\rightarrow -i\tau$ and the imaginary time $\tau$ is compactified on a circle with radius $\beta=1/T$, where $T$ is the temperature. Chemical potentials $\mu_{n,p}$ for neutrons and protons are introduced by adding a term $\Delta S_{\stext E}=-\sum_{i=n,p}\frac{\mu_i} T\int d^3x\;\psi_i^\dagger\psi_i$ to the Euclidean action. 

As a first step the model is studied in the mean-field approximation. The nucleons contribute only quadratically and are integrated out in the path-integral formalism, which leaves us with a determinant. Next, the bosonic fields are replaced by space-time independent background fields. A possible pion condensate is not considered; hence the pion mean field vanishes. Rotational invariance implies that the mean field values of the spatial components $\omega_i$ and $\boldsymbol\rho_i$ vanish. The only non-vanishing mean fields are $\sigma$, $\omega_0$ and $\rho_0^3$. By a slight abuse of notation, their mean-field values are in the following denoted by the same symbols. The effect of the mean fields is to generate an in-medium nucleon mass, $M_N$, as well as shifted effective chemical potentials for the nucleons:
\begin{align}
	M_N=g\sigma\,,\quad \mu_{n,p}^{\stext{eff}}=\mu_{n,p}-g_\omega\omega_0\pm g_\rho \rho_0^3\,.
\end{align}
The $\rho_0^3$-component introduces an isospin asymmetry. For symmetric nuclear matter, $\rho_0^3=0$.

The full mean-field potential,
\begin{align}
	U^{\stext{MF}}&=U_{\stext F}(T,\mu_{n,p},\sigma,\omega_0,\rho_0^3)+U_{\stext B}(\sigma,\omega_0,\rho_0^3)\,,
\end{align}
is split into a fermionic part (which stems from the fermion determinant) and a bosonic potential (which is independent of temperature and chemical potentials):
\begin{align}\label{eq:MF_potential}
	\begin{aligned}
		U_{\stext F}&=-2\sum_{\substack{i=n,p}}\int\frac{d^3p}{(2\pi)^3}\bigg[E_N \\
		&\hspace{2cm} +\frac{p^2}{3E_N}\Big(n_{\stext F}(\mu_i^{\stext{eff}}) + n_{\stext F}(-\mu_i^{\stext{eff}})\Big)\bigg]\,, \\
		U_{\stext B}&= \sum_{i=1}^{N_{\stext{max}}}\frac{a_n}{n!}(\chi-\chi_0)^n-m_\pi^2f_\pi(\sigma-f_\pi)\\
		&\quad-\frac 12m_v^2\Big(\omega_0^2+(\rho_0^3)^2\Big)\,,
	\end{aligned}
\end{align}
where \mbox{$E_N=\sqrt{p^2+M_N^2}$\,}, and the Fermi distribution is given by \mbox{$n_{\stext F}(\mu)=(\e^{(E_N-\mu)/T}+1)^{-1}$}. The $\chi$-dependent part of the  bosonic potential $U_{\stext B}$ is expanded around its vacuum value at $T=0$ and $\mu=0$, namely \mbox{$\chi_0=\frac 12f_\pi^2$}. The first term in the fermionic potential $U_{\stext F}$ proportional to the integral over $E_N$ has a quartic divergence. This integral can be computed in dimensional regularization as demonstrated in Ref.~\cite{skokov2010vacuum}:
\begin{align}
	\Delta U^{\stext{MF}} = -\frac{M_N^4}{8\pi^2}\log\frac{M_N^2}{\lambda^2}\,,
\label{eq:log}
\end{align}
where $\lambda$ is a renormalization scale. After making the replacement \mbox{$M_N^2\rightarrow 2g^2\chi$}, this term is added as a non-analytic contribution in $\chi$ to the bosonic potential $U_{\stext B}$. Since the microscopic potential $\mathcal U(\sigma,\boldsymbol\pi)$ is not known a priori, it is not possible to compute the bosonic contribution $U_{\stext B}$ to the mean-field potential completely from first principles. Instead, the parameters $a_n$ are fitted to reproduce empirical properties of symmetric nuclear matter and pure neutron matter, as will be shown in the following.

For given temperature $T$ and chemical potentials $\mu_{n,p}$, the mean-field potential is minimized as a function of its parameters $\sigma$, $\omega_0$ and $\rho_0^3$. The corresponding mean field equations are
\begin{align}\label{eq:MF}
	\begin{gathered}
		\frac{\partial U_{\stext B}}{\partial\sigma}=-g\,n_s\,,\quad g_\omega\,\omega_{\stext{0,min}}=G_\omega(n_p+n_n)\,,\\
		g_\rho\,\rho^3_{\stext{0,min}}=G_\rho(n_p-n_n)\,,
	\end{gathered}
\end{align}
with neutron-, proton- and scalar-density functionals defined as follows:
\begin{align}\label{eq:densities}
	\begin{gathered}
		n_i=2\int\frac{d^3p}{(2\pi)^3}\Big[n_{\stext F}(\mu_i^{\stext{eff}})+n_{\stext F}(-\mu_i^{\stext{eff}})\Big]\,,\;\; i=n,p\,, \\
		n_s=2\sum_{i=n,p}\int\frac{d^3p}{(2\pi)^3}\frac m{E_N}\Big[n_{\stext F}(\mu_i^{\stext{eff}})+n_{\stext F}(-\mu_i^{\stext{eff}})\Big]\,.
	\end{gathered}
\end{align}
The values of the fields at the minimum of the potential are denoted as $\bar\sigma(T,\mu_{n,p})$, $\bar\omega_0(T,\mu_{n,p})$ and $\bar\rho_0^3(T,\mu_{n,p})$, respectively. The potential $U^{\stext{MF}}$ evaluated at the minimum equals the grand canonical partition function,
\begin{align}
	\Omega(T,\mu_{n,p})&\equiv U^{\stext{MF}}(T,\mu_{n,p},\bar\sigma,\,\bar\omega_0,\,\bar\rho_0^3)\,.
\end{align}
Pressure $p$, proton and neutron number densities $n_{i=n,p}$, entropy density $s$, and energy density $\epsilon$ are determined from the standard thermodynamical relations:
\begin{align}\label{eq:thermodynamics}
	\begin{gathered}
		p=-\Omega(T,\mu_{n,p})~,~~~\; n_i=-\frac{\partial \Omega(T,\mu_{n,p})}{\partial\mu_i}\,,\\
		s=-\frac{\partial \Omega(T,\mu_{n,p})}{\partial T}~,~~~\;\epsilon=-p+Ts+\sum_{i=n,p}\mu_i n_i\,.
	\end{gathered}
\end{align}
Next, the parameters of the model are determined, following Ref.~\cite{floerchinger2012chemical}. First, the scalar-pseudoscalar coupling $g$ is fixed to reproduce the correct nucleon mass in vacuum at $T=0$ and $\mu=0$, which gives $g=939\text{ MeV}/f_\pi=10.1$. At vanishing temperature and chemical potential, the minimum must be located at $\sigma=f_\pi$ with vanishing pressure $p_{\stext{vac}}$, and therefore the mean-field potential must satisfy
\begin{align}
	U^{\stext{MF}}\Big|_{\sigma=f_\pi}=-p_{\stext{vac}}=0\,,\quad \left.\frac{\partial U^{\stext{MF}}}{\partial\sigma}\right|_{\sigma=f_\pi}=0\,.
\end{align}
The masses of the physical pion field and of the scalar field $\sigma$ are:
\begin{align}
	\left.\frac{\partial U^{\stext{MF}}}{\partial\chi}\right|_{\chi_0}=m_\pi^2\,,\;\, \left.\left(\frac{\partial U^{\stext{MF}}}{\partial\chi}+2\chi\frac{\partial^2U^{\stext{MF}}}{\partial\chi^2}\right)\right|_{\chi_0}=m_\sigma^2\,.
\end{align}
The mass $m_\sigma$ is a free parameter so far. The ansatz \eqref{eq:MF_potential} for the mean-field potential obeying the necessary constraints is given by:
\begin{align}\label{eq:MF_potential_2}
	\begin{gathered}
		U^{\stext{MF}}=-2\sum_{\substack{i=n,p}}\int\frac{d^3p}{(2\pi)^3}\frac{p^2}{3E_N}\left[n_{\stext F}(\mu_i^{\stext{eff}})+n_{\stext F}(-\mu_i^{\stext{eff}})\right] \\
		+ \left[m_\pi^2+\frac{g^4}{4\pi^2}f_\pi^2\left(1+2\log\frac{f_\pi^2}{2\chi}\right)\right]\,(\chi-\chi_0) \\
		+ \frac 12\left[\frac{m_\sigma^2-m_\pi^2}{f_\pi^2}+\frac{g^4}{2\pi^2}\left(3+2\log\frac{f_\pi^2}{2\chi}\right)\right]\,(\chi-\chi_0)^2 \\
		+ \frac{g^4}{8\pi^2}f_\pi^4\log\frac{f_\pi^2}{2\chi} + \sum_{n=3}^{N_{\stext{max}}}\frac{a_n}{n!}(\chi-\chi_0)^n \\
		- m_\pi^2f_\pi(\sigma-f_\pi)-\frac 12m_v^2\Big(\omega_0^2+(\rho_0^3)^2\Big)\,.
	\end{gathered}
\end{align}
We observe that the dependence on the renormalization scale $\lambda$ has dropped out as it should \cite{skokov2010vacuum}. Moreover, the mean-field potential $U^{\stext{MF}}$ is finite for $\chi\rightarrow 0$.

\begin{figure}
	\centering
	\begin{overpic}[width=0.4\textwidth]{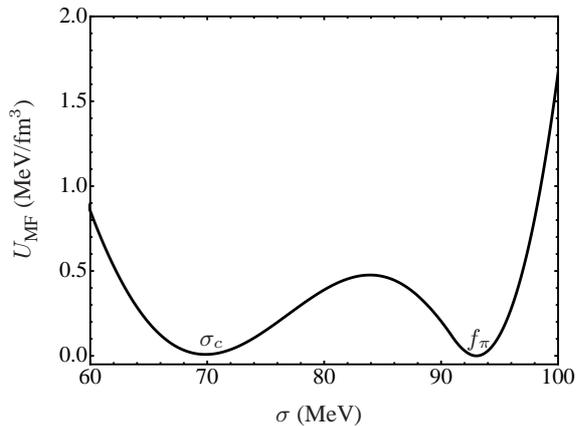}
		\put(0,30){\begin{rotate}{90}\small $U_{\stext{MF}}$ (MeV/fm${}^3$)\end{rotate}}
		\put(45,0){\small $\sigma$ (MeV)}
		\put(31,14){\small $\sigma_c$}
		\put(80,14){\small $f_\pi$}
	\end{overpic}
	\caption{Mean-field potential of symmetric nuclear matter as function of $\sigma$ at vanishing temperature for $\mu=\mu_c = 923$ MeV.}
	\label{fig:MF_potential}
\end{figure}
In the following, we choose \mbox{$N_{\stext{max}}=4$}, which leaves us with five free parameters: $a_3$, $a_4$, $G_\omega = g_\omega^2/m_v^2$, $G_\rho=g_\rho^2/m_v^2$ and $m_\sigma$. Some of these parameters are fixed by empirical properties of symmetric nuclear matter. In the isospin-symmetric case we have \mbox{$n_p=n_n$} and $\rho_0^3$ vanishes identically, as seen from the mean-field equation~\eqref{eq:MF}. Hence, $G_\rho$ does not enter the calculations in this case and there is only a single chemical potential, \mbox{$\mu\equiv\mu_p=\mu_n$}. 

At vanishing temperature, the liquid-gas phase transition sets in at a critical chemical potential equal to the difference between nucleon mass and binding energy, \mbox{$\mu_c=M_N-B = 923\,\text{MeV}$} with $B = 16$ MeV. As shown in Fig.~\ref{fig:MF_potential}, the potential has two degenerate minima, where the one at \mbox{$\sigma=f_\pi$} corresponds to the vacuum, while the minimum at \mbox{$\sigma_c=69.8\,\text{MeV}$} corresponds to nuclear matter in its ground state. Let $\omega_{0,c}$ be the expectation value of \mbox{$\omega_0$ for $T=0$} and \mbox{$\mu=\mu_c$} at $\sigma=\sigma_c$. The effective chemical potential at $\mu_c$ is given by \mbox{$\mu_c-g_\omega\,\omega_{0,c}$}. It is equal to the Landau mass, $m_{\stext L}$, the effective mass associated with a nucleon quasi-particle excitation at the Fermi surface, i.e., \mbox{$m_{\stext L}=\sqrt{p_{\stext F}^2+(g\sigma_c)^2}$}, where \mbox{$p_{\stext F}=265\,\text{MeV}$} is the Fermi momentum at nuclear saturation density $n_0$ and \mbox{$g\sigma_c$} is the dynamical in-medium nucleon mass. For symmetric nuclear matter at $\mu=\mu_c$, the neutron and proton densities are equal and sum up to the nuclear saturation density $n_0$. With Eq.~\eqref{eq:MF} we find
\begin{align}
	m_{\stext L}=\mu_c-g_\omega\,\omega_{0,c}=\mu_c-G_\omega n_0\,~.
\end{align}
Given the Landau mass \mbox{$m_{\stext L}\simeq 0.8\,M_{\stext N}$} fixes 
\begin{equation}
G_\omega=g_\omega^2/m_v^2=5.71\text{ fm}^2 \nonumber
\end{equation}
at mean-field level. As already mentioned, this is the only relevant combination of $g_\omega$ and $m_v$ as long as $\omega_0$ is kept as a non-fluctuating background field. It is the coupling strength of a corresponding nucleon contact interaction, $G_\omega(\psi^\dagger\psi)^2$.

As mentioned, the condition of a first-order phase transition at \mbox{$\mu=\mu_c$} implies two degenerate minima of the mean-field potential $U^{\stext{MF}}$, located at $\sigma =\sigma_c$ and $\sigma = f_\pi$. The mean-field potential must therefore satisfy the constraints
\begin{align}
	\begin{gathered}
		\left.\frac{\partial U^{\stext{MF}}}{\partial\sigma}\right|_{T=0,\mu_c,\sigma_c,\,\omega_{0,c}} = 0\,, \\
		U^{\stext{MF}}\big|_{T=0,\mu_c,\sigma_c,\,\omega_{0,c}}=U^{\stext{MF}}\big|_{T=0,\mu_c,f_\pi,0}\,~.
	\end{gathered}
\end{align}
These conditions allow us to solve for $a_3$ and $a_4$ as a function of $m_\sigma$, such that only $m_\sigma$ remains as a free parameter for symmetric nuclear matter. At this point it is important to note that the sigma mass is not to be identified with the complex pole which appears in pion-pion scattering in the \mbox{$I=0$} $s$-wave channel at \mbox{$\sqrt s\simeq (500-i\,300)\text{ MeV}$} \cite{colangelo2001pipi,garcia-martin2007experimental}. Instead, the $\sigma$ boson in our model parametrizes part of the short-distance interaction. Its value is determined in order to achieve good agreement with the compression modulus \mbox{$K=9n(dn/d\mu)^{-1}$} and the nuclear surface tension \mbox{$\Sigma=\int_{\sigma_0}^{f_\pi}d\sigma\,\sqrt{2U}$} \cite{coleman1977fate}. A good choice is\footnote{
	Note that these parameters differ from our earlier study \cite{drews2013thermodynamic}, where the
term, Eq.\,(\ref{eq:log}), was not explicitly included in the mean-field potential.
}
\begin{align}
	\begin{gathered}
	m_\sigma=880\text{ MeV}\,,\quad a_3=6.87\cdot10^{-2}\text{ MeV}^{-2}\,, \\
	a_4=2.05\cdot 10^{-4}\text{ MeV}^{-4}\,,
	\end{gathered}
\end{align}
which gives \mbox{$K=293\text{ MeV}$} and \mbox{$\Sigma=1.1\text{ MeV}/\text{fm}^{2}$}, as compared to the empirical values \mbox{$K=240\pm30\text{ MeV}$} and \mbox{$\Sigma=1.1\text{ MeV}/\text{fm}^{2}$}, respectively. The behavior of the liquid-gas transition and in particular its critical end-point are sensitive to $K$ and $\Sigma$. It is therefore important to satisfy the empirical constraints.

Only $G_\rho$ remains as a free parameter. Again, only the combination \mbox{$G_\rho=g_\rho^2/m_v^2$} enters the equations, corresponding to the strength of an isovector-vector four-nucleon interaction \mbox{$G_\rho(\psi^\dagger\boldsymbol\tau\psi)^2$}. The parameter $G_\rho$ is fitted to reproduce the symmetry energy $E_{\stext{sym}}$ at nuclear saturation density $n_0$. The symmetry energy $S(n)$ is defined as the difference between the energy per particle of pure neutron matter and symmetric nuclear matter at a given density $n$: 
\begin{align}\label{eq:SL}
	\begin{gathered}
		\frac EA(n,x)=\frac EA(n,0.5) + S(n)\,(1-2x)^2+\ldots\,,\\
		S(n)=E_{\stext{sym}}+\frac L3(n-n_0)+\ldots\,,
	\end{gathered}		
\end{align}
where the proton fraction $x=Z/A=n_p/(n_p+n_n)$ is a measure of asymmetry. The $L$-parameter is related to the slope of the symmetry energy as a function of density $n$ around nuclear saturation density. The symmetry energy and the $L$-parameter can be inferred from measurements of neutron skin thickness, heavy ion collisions, dipole polarizabilities, giant and pygmy dipole resonance energies, as well as from fitting nuclear masses. A combined analysis gives values in the range \mbox{$29\text{\,MeV}\lesssim E_{\stext{sym}}\lesssim33\text{\,MeV}$} and \mbox{$40\text{\,MeV}\lesssim L \lesssim 62\text{\,MeV}$} \cite{tsang2012constraints,lattimer2013constraining,lattimer2014constraints}.  Reproducing the symmetry energy value \mbox{$E_{\stext{sym}}=32\text{\,MeV}$} fixes $G_\rho=1.07\text{ fm}^2$. All model parameters are now determined. Next, we go beyond the mean-field analysis and study the influence of fluctuations.

\section{Inclusion of fluctuations: \\functional renormalization group}\label{fluctuations}
Fluctuations beyond the mean-field approximation in the ChNM model are included using the framework of the functional renormalization group \cite{drews2013thermodynamic}. A proper treatment of fluctuations was shown to improve the agreement with calculations performed within in-medium chiral effective field theory \cite{fiorilla2012chiral}. The functional renormalization group (FRG) is a method to compute the full quantum effective action $\Gamma_{\stext{eff}}$ from a given initial action defined at an ultraviolet cutoff scale $\Lambda$ \cite{berges2002non-perturbative,pawlowski2007aspects,gies2012introduction,delamotte2012introduction,braun2012fermion}. To this end, an effective action $\Gamma_k$ is introduced, which depends on a renormalization scale $k$. The flow of the effective action is determined in such a way that it interpolates between the ultraviolet action at the scale $\Lambda$ and the full quantum effective action \mbox{$\Gamma_{\stext{eff}}=\Gamma_{k=0}$} in the infrared limit $k\rightarrow 0$. The flow of $\Gamma_k$ as a function of $k$ is given by a functional differential equation, Wetterich's flow equation \cite{wetterich1993exact}:
\begin{align}
	\begin{aligned}
		k\,\frac{\partial\Gamma_k}{\partial k}=
		\begin{aligned}
			\vspace{1cm}
			\includegraphics[width=0.08\textwidth]{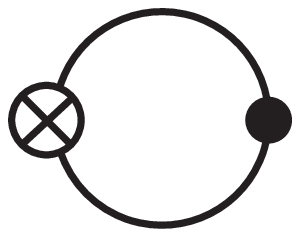}
		\end{aligned} \vspace{-1cm}=\frac 12 \operatorname{Tr}\frac{k\,\frac{\partial R_k}{\partial k}}{\Gamma_k^{(2)}+R_k}\,,
	\end{aligned}
\end{align}
where $\Gamma^{(2)}_k$ is the second functional derivative of the effective action with respect to the fields, and Tr stands for Dirac and isospin traces as well as integration over loop momentum. Pictorially, the line with the dot represents the full propagator of the fluctuating degrees of freedom (pions, $\sigma$ field and nucleons), while the cross symbolizes the insertion of a regulator function $k\partial R_k/\partial k$. The regulator $R_k$ ensures that the flow equation is IR-finite. The fluctuations contributing to the flow equation at a scale $k$ have momenta peaked around $k$. The optimized Litim-cutoff \cite{litim2006non-perturbative,blaizot2007perturbation} is chosen, namely
\begin{align}\label{eq:cutoff}
	R_k(\boldsymbol p^2)=(k^2-\boldsymbol p^2)\,\theta(k^2-\boldsymbol p^2)\,,
\end{align}
where $\boldsymbol p$ is the three-momentum. 

The mass $m_v$ associated with the $\omega$ and  $\rho$ bosons correspond to the inverse range of isoscalar and isovector short-distance NN interactions. This mass is large compared to the relevant low-energy scales. Therefore, $\omega$ and $\rho$ fluctuations are suppressed and these "frozen" degrees of freedom are treated as background fields in the mean-field approximation as before. In contrast, the fluctuations of the pions and (in order to maintain chiral symmetry) also of the $\sigma$ are included, as well as important particle-hole excitations of the nucleons around the Fermi surface. 

For the treatment of the thermodynamics with inclusion of fluctuations it is useful to compute the flow of the difference between the effective action at given values of temperature and chemical potential, $\Gamma_k(T,\mu)$, as compared to the potential at the liquid-gas phase transition point at zero temperature, $\Gamma_k(0,\mu_c)$, at which nuclear matter is in equilibrium. In analogy to Ref.~\cite{litim2006non-perturbative}, we study the flow of the difference
\begin{align}
	\bar\Gamma_k=\Gamma_k(T,\mu)-\Gamma_k(0,\mu_c)\,.
\end{align}
The $k$-dependence of $\Gamma_k$ is given by
\begin{align}
	\begin{aligned}
		\frac{k\,\partial\bar \Gamma_k}{\partial k}(T,\mu)&=
		\begin{aligned}
			\hspace{-.1cm}
			\vspace{1cm}
			\includegraphics[width=0.08\textwidth]{graphicsEPS/wetterich_fermion.eps}
		\end{aligned} \vspace{-1cm}\Bigg|_{T,\mu}-
		\begin{aligned}
			\hspace{-.1cm}
			\vspace{1cm}
			\includegraphics[width=0.08\textwidth]{graphicsEPS/wetterich_fermion.eps}
		\end{aligned} \Bigg|_{\begin{subarray}{l} T=0 \\ \mu=\mu_c \end{subarray}}.
	\end{aligned}
\end{align}
The effective action is treated in leading order of the derivative expansion, i.e., operators with higher powers in derivatives are not included. We work in the local potential approximation, which means that a  possible anomalous dimension (a $Z$-factor) or the so-called $Y$-term, which includes a derivative coupling together with higher powers in the fields, \mbox{$Y(\phi)\,(\partial\phi)^2$}, are not considered. Moreover, the running of the Yukawa couplings is ignored. With these simplifications, the effective action is written as
\begin{align}
	\begin{aligned}
	&\Gamma_k=\int d^4x\;\bigg\{\bar\psi i\slashed\partial\psi+\frac 12\partial_\mu\sigma\,\partial^\mu\sigma+\frac 12\partial_\mu\boldsymbol\pi\cdot\partial^\mu\boldsymbol\pi \\
	&-\bar\psi\Big[g(\sigma+i\gamma_5\,\boldsymbol\tau\cdot\boldsymbol\pi)+\gamma_0(g_\omega\, \omega_0+g_\rho\,\rho_0^3\tau^3)\Big]\psi-U_k\bigg\}\,.
	\end{aligned}
\end{align}
As mentioned, the vector fields $\omega_0$ and $\rho_0^3$ appear here only as mean fields. The complete $k$-dependence is in the effective potential $U_k$. In analogy to the mean-field potential \eqref{eq:MF_potential_2}, the effective potential has a chirally symmetric piece, $U^{(\chi)}$, the explicit chiral symmetry breaking term and the mass terms of the vector bosons:
\begin{align}\label{eq:decomposition}
	U_k=U_k^{(\chi)}-m_\pi^2f_\pi(\sigma-f_\pi)-\frac 12m_v^2\Big(\omega_0^2+(\rho_0^3)^2\Big)\,.
\end{align}
The second derivative $\Gamma^{(2)}$ is computed and the Dirac and isospin trace is performed.
Due to the choice of the optimized regulator \eqref{eq:cutoff}, the only momentum dependence comes in through a step function, and the momentum integral can be performed trivially. The remaining flow equations depend only on the chirally invariant field $\chi$.

The flow of the subtracted chirally symmetric potential \mbox{$\bar U_k^{(\chi)}=U_k^{(\chi)}(T,\mu)-U^{(\chi)}_k(0,\mu_c)$} is computed from the equation
\begin{gather}\label{eq:flow_equation}
	\frac{\partial\bar U_k^{(\chi)}(T,\mu)}{\partial k}=f_k(T,\mu)-f_k(0,\mu_c)\,,
\end{gather}
with
\begin{multline}
	f_k(T,\mu)=\frac {k^4}{12\pi^2} \bigg\{3\cdot\frac{1+2n_{\stext B}(E_\pi)}{E_\pi}+\frac {1+2n_{\stext B}(E_\sigma)}{E_\sigma} \\
	-4 \sum_{i=n,p}\frac{1-n_{\stext F}\left(E_N,\mu^{\stext{eff}}_i\right)-n_{\stext F}\left(E_N,-\mu^{\stext{eff}}_i\right)}{E_N}\bigg\}\,.
\end{multline}
Here,
\begin{align}\label{eq:mpi}
	\begin{gathered}
		E_N^2=k^2+2g^2\chi\,, \\
		E_\pi^2=k^2+\frac{\partial U_k}{\partial\chi}\,,\quad E_\sigma^2=k^2+\frac{\partial U_k}{\partial\chi}+2\chi\frac{\partial^2U_k}{\partial\chi^2}\,, \\
		n_{\stext B}(E)=\frac 1{\e^{E/T}-1}\,,~~\text{ and }~˝\, n_{\stext F}(E,\mu)=\frac 1{\e^{(E-\mu)/T}+1}\,.
	\end{gathered}
\end{align}
So far $\omega_0$ and $\rho_0^3$ are constant background fields, to be determined self-consistently in such a way that the effective potential at $k=0$ is minimized as a function of $\omega_0$ and $\rho_0^3$. Instead, it is possible to introduce a $k$- and $\chi$-dependence for these fields, such that the effective potential $U_k$ is minimized at each scale $k$ for each $\chi$. From Eq.~\eqref{eq:decomposition} follow the two gap equations for $\omega_0(k,\chi)$ and $\rho_0^3(k,\chi)$:
\begin{align}
	\begin{gathered}
		\frac\partial{\partial y}\Big[U_k^{(\chi)}\Big(y,\rho_0^3(k,\chi)\Big)-\frac 12m_v^2y^2\Big]\Big|_{y=\omega_0(k,\chi)}=0\,, \\
		\frac\partial{\partial y}\Big[U_k^{(\chi)}\Big(\omega_0(k,\chi),y\Big)-\frac 12m_v^2y^2\Big]\Big|_{y=\rho^3_0(k,\chi)}=0\,.
	\end{gathered}
\end{align}
With the help of the flow equation \eqref{eq:flow_equation} it is possible to rewrite these gap equations in the following form:
\begin{gather}\label{eq:flow_equation_2}
	\begin{gathered}
	g_\omega\,\omega_0(k,\chi) = 
\frac {G_\omega}{3\pi^2}\int_k^\Lambda dp \; \frac{p^4}{E_N} \hspace{3.3cm} \\
	\quad\cdot\sum_{r=\pm1}\frac\partial{\partial \mu}\Big[n_{\stext F}\big(E_N,r\mu^{\stext{eff}}_p(k,\chi)\big)+n_{\stext F}\big(E_N,r\mu^{\stext{eff}}_n(k,\chi)\big)\Big] \,, \\
	g_\rho\,\rho^3_0(k,\chi) = \frac {G_\rho}{3\pi^2}\int_k^\Lambda dp \; \frac{p^4}{E_N} \hspace{3.3cm} \\
	\quad\cdot\sum_{r=\pm1}\frac\partial{\partial \mu}\Big[n_{\stext F}\big(E_N,r\mu^{\stext{eff}}_p(k,\chi)\big)-n_{\stext F}\big(E_N,r\mu^{\stext{eff}}_n(k,\chi)\big)\Big] \,,
	\end{gathered}
\end{gather}
where the effective chemical potentials now depend on $k$ and $\chi$ according to
\begin{align}
	\mu^{\stext{eff}}_{n,p}(k,\chi)=\mu_{n,p}-g_\omega\,\omega_0(k,\chi) \pm g_\rho\,\rho_0^3(k,\chi)\,.
\end{align}
These equations can be considered as generalizations of the mean field equations \eqref{eq:MF} in the context of the functional renormalization group. After an integration by parts, the gap equations can be brought into a form similar to Eq.~\eqref{eq:densities}, where the momenta of the nucleons contributing to the mean fields $\omega_0(k,\chi)$ and $\rho_0^3(k,\chi)$ at a certain step are restricted to the range \mbox{$k\le p \le\Lambda$}, as is clear from the integral boundaries of Eq.~\eqref{eq:flow_equation_2}. In addition, boundary terms from the integration by parts appear, which however vanish in the limit $k\rightarrow 0$ and for large cutoffs $\Lambda$. In this way, it is possible to show that the flow equations reproduce the mean field results if bosonic loops are ignored.

Finally, the ultraviolet potential is fixed in such a way that for \mbox{$T=0$} and \mbox{$\mu=\mu_c$} the mean field potential \eqref{eq:MF_potential} is reproduced. This guarantees that nuclear matter is described accurately. Moreover, the pion mass and pion decay constant, both determined from the behavior of the potential at its minimum, are correctly kept.

As explained in Ref.~\cite{drews2013thermodynamic}, the input parameters have to be readjusted in order to reproduce the correct nuclear saturation density, because the dependence of the minimum of the effective potential on the chemical potentials is influenced by the fluctuations. Again, the coupling $G_\rho$ is fixed  to reproduce a symmetry energy of 
\mbox{$E_{\stext{sym}}=32\text{ Mev}$}. The updated parameters of the model with inclusion of fluctuations are
\begin{align}
	\begin{gathered}
		G_\omega=\frac{g_\omega^2}{m_v^2}=4.04\text{ fm}^2\,,\quad G_\rho=\frac{g_\rho^2}{m_v^2}=1.12\text{ fm}^2\,, \\
		m_\sigma=770\text{ MeV}\,,\quad a_3=5.55\cdot 10^{-3}\text{ MeV}^{-2}\,,\\
		a_4=8.38\cdot 10^{-5}\text{ MeV}^{-4}\,.
	\end{gathered}
\end{align}
We choose an ultraviolet cutoff \mbox{$\Lambda=1.4\text{ GeV}$}, slightly above the chiral scale $4\pi f_\pi$ below which the hadronic effective Lagrangian ${\cal L}_{\stext{ChNM}}$ is applicable. For given temperature $T$ and chemical potentials $\mu_n$ and $\mu_p$, the full set of flow equations \eqref{eq:flow_equation} and \eqref{eq:flow_equation_2} is solved self-consistently using the grid method \cite{adams1995solving}. The potential is expanded as a function of $\chi$ around grid points and then matched continuously between any two adjacent grid points. In this way, the potential is not Taylor expanded but kept as a general function of $\chi$. Upon expansion of the effective potential around its absolute minimum, $n$-point interactions involving pions and sigma fields are generated in the effective action. In the context of a generalized linear sigma model, multi-pion and $\sigma$ exchange interactions are thus incorporated to all orders. In contrast to (perturbative) in-medium chiral effective field theory based on a non-linear sigma model, the $n$-point correlators in the effective action are now computed in a fully non-perturbative fashion.
 
The grand canonical potential $\Omega$ is finally obtained by evaluating the effective potential as a function of $\sigma$ at its absolute minimum. From this grand canonical potential all thermodynamic properties, such as pressure, energy density and entropy, can be derived as in Eq.~\eqref{eq:thermodynamics}.

\section{From asymmetric nuclear matter to neutron matter}\label{sec:asymmetric}
\begin{figure}
	\centering
	\begin{overpic}[width=0.45\textwidth]{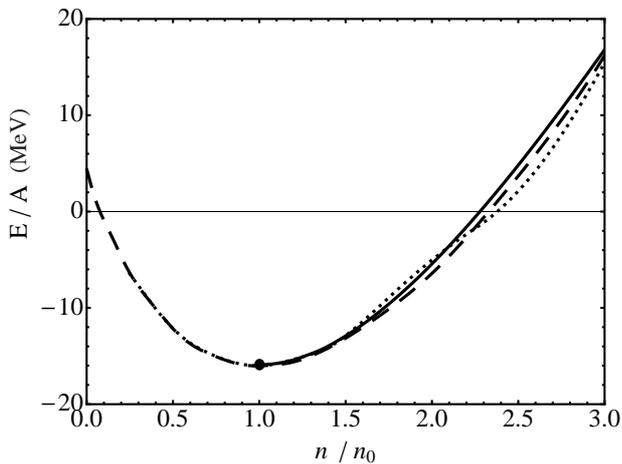}
		\put(3,49){\begin{rotate}{90} (MeV) \end{rotate}}
	\end{overpic}
	\vspace{-0.2cm}
	\caption{Energy per particle of symmetric nuclear matter at $T = 0$ as a function of baryon density (in units of $n_0 = 0.16$ fm$^{-3}$) computed in the FRG-ChNM model (solid line), as compared to the Akmal-Pandharipande-Ravenhall EoS (dotted, \cite{akmal1998equation}), and a QMC computation (dashed, \cite{armani2011recent})}
	\label{fig:e_a_sym}
\end{figure}

Consider first symmetric nuclear matter at small temperatures, which exhibits a first-order transition between a gas phase and a nuclear-liquid phase. The absolute minimum of the energy per particle is located at saturation density $n_0$ and equals the binding energy, $E/A = -B = -16\text{ MeV}$. In Fig.~\ref{fig:e_a_sym}, the energy per particle at vanishing temperature $T$ is shown as a function of density in the region of stability\footnote{Continuations of curves often extended into the unstable region $n < n_0$ are omitted here.}. For comparison, the equation of state by Akmal, Pandharipande, and Ravenhall (APR) \cite{akmal1998equation} based on phenomenological two- and three-body potentials is shown, as well as a Quantum Monte Carlo (QMC) computation \cite{armani2011recent}. All results are in very good mutual agreement, even up to densities as high as three times nuclear saturation density. 

\begin{figure}
	\centering
	\begin{overpic}[width=0.45\textwidth]{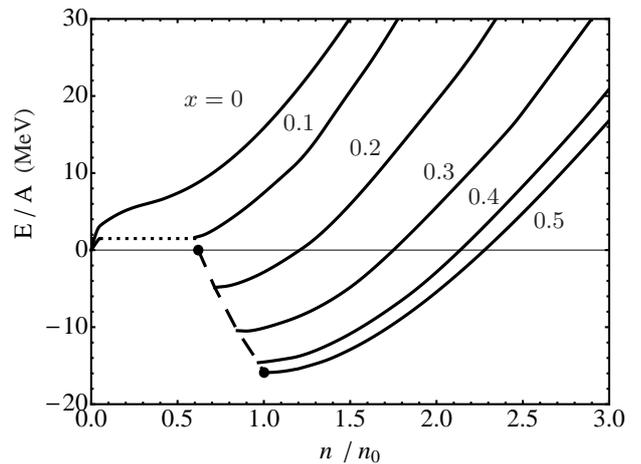}
		\put(3,49){\begin{rotate}{90} (MeV) \end{rotate}}
		\put(85,40){$0.5$}
		\put(74.3,44){$0.4$}
		\put(67,48){$0.3$}
		\put(55,52){$0.2$}
		\put(44.2,56){$0.1$}
		\put(28,60){$x=0$}
	\end{overpic}
	\vspace{-0.2cm}
	\caption{The equation of state for different proton fractions $x = Z/A$ at vanishing temperature. The dashed curve denotes the absolute minimum of the energy per particle for each asymmetry $x$. Matter in the region to the left of the dashed curve is unstable. The dotted line results from a Maxwell construction.\label{fig:e_a}}
\end{figure}

\begin{figure}
	\centering
	\begin{overpic}[width=0.45\textwidth]{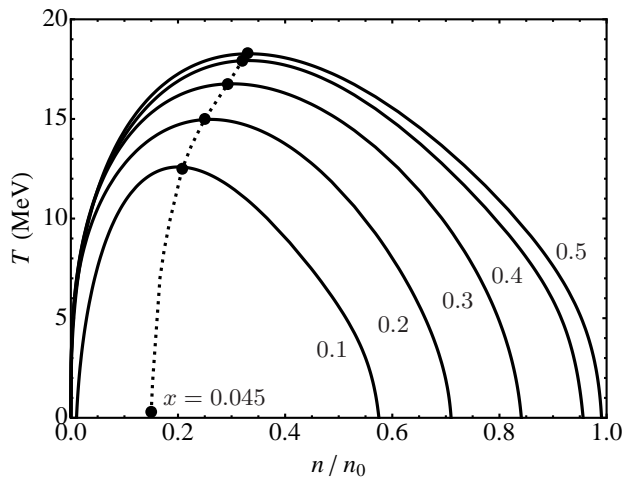}
		\put(89,36){$0.5$}
		\put(78.5,32){$0.4$}
		\put(70.5,28){$0.3$}
		\put(60,24){$0.2$}
		\put(50,20){$0.1$}
		\put(25,12){$x=0.045$}
	\end{overpic}
	\vspace{-0.2cm}
	\caption{The liquid-gas coexistence regions for different proton fractions $x = Z/A$. The trajectory of the critical point is shown as the dotted line.\label{fig:coexistence}}
\end{figure}

In Fig.~\ref{fig:e_a} the energy per particle at $T = 0$ as a function of density is shown for different proton fractions, $x = Z/A$, from symmetric nuclear matter (\mbox{$x=0.5$}) to pure neutron matter (\mbox{$x=0$}). As the proton fraction $x$ is lowered, the energy per particle increases, until for \mbox{$x\simeq0.11$} the energy per particle vanishes at the minimum (the upper endpoint of the dashed curve in Fig.~\ref{fig:e_a}). For even smaller values of $x$ the absolute minimum occurs at zero density and nuclear matter is no longer self-bound. There is still a remnant of the first-order phase transition, and the density is still discontinuous as a function of the chemical potential. However, the coexistence region extends no longer down to vanishing density \mbox{$n=0$}. In Fig.~\ref{fig:coexistence} the coexistence regions in a temperature/density-plot are shown for different proton fractions. For instance, for $x=0.1$, the coexistence region starts at non-vanishing density determined by a Maxwell construction from the energy per particle, and depicted as the dotted line in Fig.~\ref{fig:e_a} for \mbox{$x=0.1$}. Finally, for $x$ smaller than a critical value of \mbox{$x=0.045$} the energy per particle rises monotonously as a function of density. There is no longer a second minimum and the coexistence region vanishes altogether, as is seen in Fig.~\ref{fig:coexistence}.

As the temperature increases, the phase coexistence region melts until it disappears at a certain $x$-dependent critical temperature characterized by a second-order critical endpoint. From the behavior of the coexistence regions one can read off the critical endpoint for symmetric matter, which is located at a temperature \mbox{$T=18.3\text{ MeV}$} and a density \mbox{$n=0.053\text{ fm}^{-3}$}. These values are in excellent agreement with analyses of compound nuclear reactions and multifragmentation experiments, which give critical temperatures of \mbox{$T=17.9\pm0.4\text{ MeV}$} and critical densities \mbox{$n=0.06\pm0.01\text{ fm}^{-3}$} \cite{karnaukhov2008critical,elliott2013determination}. The 
trajectory of the critical endpoint as the proton fraction $x$ is varied is indicated by the dotted curve. We note that our idealized model ignores surface effects as well as Coulomb repulsion. In more realistic scenarios at low densities the effects of light clusters have to be taken into account. However, a study in the framework of relativistic mean field and microscopic quantum statistical models showed an only moderate influence of such effects on the position of the critical endpoint \cite{typel2010composition}.

\begin{figure}
	\centering
	\begin{overpic}[width=0.45\textwidth]{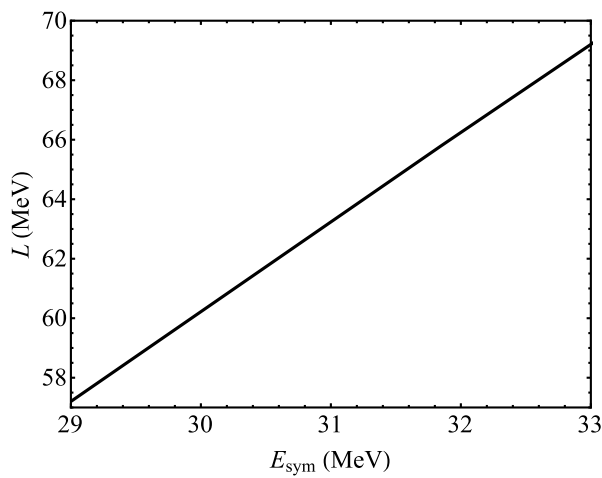}
	\end{overpic}
	\vspace{-0.2cm}
	\caption{The $L$-parameter as a function of the symmetry energy corresponding to a parameter interval \mbox{$0.91\text{ fm}^2\le G_\rho \le 1.2\text{ fm}^2$}.\label{fig:LvsEsymm}}
\end{figure}

\begin{figure}
	\centering
	\begin{overpic}[width=0.45\textwidth]{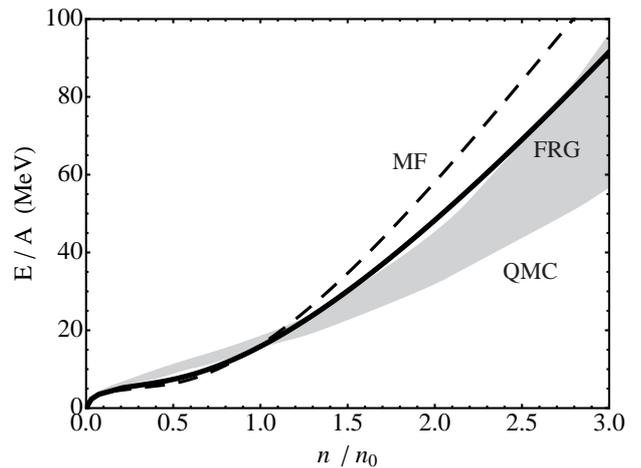}
		\put(80,32){QMC}
		\put(85,52){FRG}
		\put(62,50){MF}
	\end{overpic}
	\vspace{-0.2cm}
	\caption{The equation of state for pure neutron matter with $E_{\text{sym}}=32\text{ MeV}$ in the mean-field approximation (MF) and with fluctuation (FRG). The gray band shows QMC results \cite{gandolfi2012maximum} with different spatial and spin structures of the three-neutron interaction, and with \mbox{$32.0\text{ MeV}\le E_{\text{sym}} \le 33.7\text{ MeV}$}.} 
\label{fig:n_e}
\end{figure}

Next we study in more detail the equation of state for pure neutron matter in comparison with results of many-body computations reported in the literature. The coupling $G_\rho$ is fixed to reproduce \mbox{$E_{\stext{sym}}=32\text{ MeV}$}. The $L$ parameter corresponding to the slope of the symmetry energy as defined in Eq.~\eqref{eq:SL} is \mbox{$L=66.3\text{ MeV}$}, close to the empirical range \mbox{$40\text{ MeV}\lesssim L \lesssim 62\text{ MeV}$} \cite{lattimer2013constraining}. At this point it is instructive to examine the relationship between $L$ and the symmetry energy $E_{\stext{sym}}$ itself. As demonstrated in Fig.~\ref{fig:LvsEsymm} this dependence is approximately linear.

In Fig.~\ref{fig:n_e} the energy per particle for neutron matter calculated in the FRG-ChNM model is shown as a function of density (black line). In comparison, results obtained in a quantum Monte Carlo study with realistic two- and three-nucleon interactions \cite{gandolfi2012maximum} are shown. Note that in contrast to the mean-field approximation, the FRG treatment improves significantly the comparison with realistic ab-initio many-body calculations of E/A.

\begin{figure}
	\centering
	\begin{overpic}[width=0.45\textwidth]{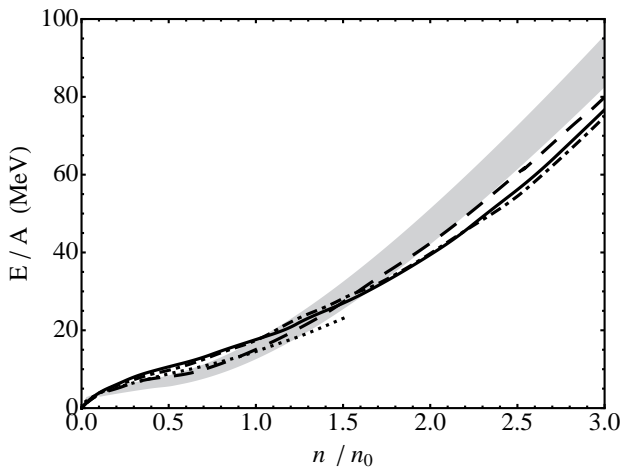}
	\end{overpic}
	\vspace{-0.2cm}
	\caption{The equation of state for pure neutron matter. The gray band are the FRG results with \mbox{$29\text{ MeV}\le E_{\text{sym}} \le 33\text{ MeV}$}. For reference, predictions from ChEFT (full line, \cite{hell2014dense}), QMC based on realistic potentials (dashed, \cite{armani2011recent}), QMC based on chiral potentials (dotted, \cite{roggero2014quantum}) as well as the Akmal-Pandharipande-Ravenhall EoS (dashed-dotted, \cite{akmal1998equation}) are shown.\label{fig:n_e_2}}
\end{figure}

Fig.~\ref{fig:n_e_2} displays a band of calculated FRG-ChNM equations of state covering a range of symmetry energies, \mbox{$29\text{ MeV}\le E_{\text{sym}} \le 33\text{ MeV}$}, corresponding to an interval \mbox{$0.91\text{ fm}^2\le G_\rho \le 1.2\text{ fm}^2$} of short-range isovector-vector couplings. Also shown for  comparison are the APR equation of state based on phenomenological potentials \cite{akmal1998equation}, results from chiral effective field theory \cite{hell2014dense}, and different QMC computations using phenomenological \cite{armani2011recent} and chiral potentials \cite{roggero2014quantum}. The equations of state obtained in the FRG-extended ChNM model agree quite well with all these results up to densities as high as \mbox{$n=3\,n_0$}.

\section{Pion mass in the nuclear medium}\label{sec:pionmass}
The pion mass can be extracted from Eq.~\eqref{eq:mpi} as \mbox{$m_\pi^2=\frac{\partial U_{k=0}}{\partial\chi}$}, where the right-hand side is evaluated at the minimum of the full effective potential at $k=0$. The density dependence of the pion mass plays an important role in low-energy pion-nuclear interactions \cite{ericson1988pions}, e.g., in the analysis of deeply-bound pionic atoms based on the s-wave pion-nucleus optical potential \cite{yamazaki1996discovery,kolomeitsev2003chiral,KKW2005}. This is an interesting test case for the role of pionic fluctuations. The threshold s-wave $\pi^-$ optical potential for isospin-symmetric nuclei is of the form
\begin{align}
V_{\stext{opt}} = -{2\pi\over m_\pi}\,b_0^{\text{eff}}\,n~,
\end{align}
where the effective scattering length,
\begin{align}
b_0^{\text{eff}}= b_0 - \left(b_0^2 + 2b_1^2\right)\langle 1/r\rangle~,
\end{align}
is dominated by the double scattering contribution involving the isospin-dependent s-wave parameter $b_1$ while the isospin-even parameter $b_0$ is small (in fact it vanishes in the chiral limit). The inverse correlation length associated with the propagating pion in the double scattering process is $\langle 1/r\rangle = 3p_F/2\pi$ for a gas of nucleons
with Fermi momentum $p_F$. Thus, the change of the pion mass in medium, $\Delta m_\pi(n) \simeq V_{\stext{opt}}(n)$,
is governed almost entirely by what the FRG scheme characterizes as pionic fluctuations, rather than being driven
by the mean-field (Hartree) term linear in the density $n$ and proportional to $b_0$. Empirically, $V_{\stext{opt}} \simeq 0.1\, m_\pi$ at $n \simeq n_0 = 0.16$ fm$^{-3}$ from the analysis of pionic atoms. 

\begin{figure}
	\centering
	\begin{overpic}[width=0.45\textwidth]{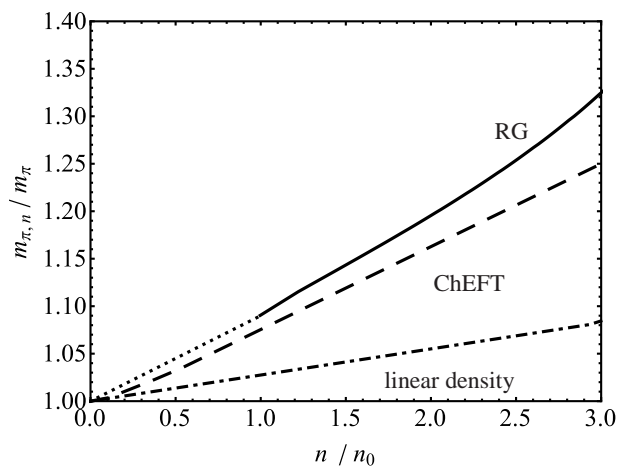}
		\put(70,30){ChEFT}
		\put(80,55){RG}
		\put(62,14){linear density}
	\end{overpic}
	\vspace{-0.2cm}
	\caption{The in-medium pion mass (normalized to the vacuum mass) as a functions of density for symmetric nuclear matter at $T=0$. Solid line: FRG-ChNM calculation, dashed line: in-medium chiral perturbation theory (ChEFT) \cite{goda2014pion}. Dash-dotted line: leading (linear) order in the density expansion.\label{fig:m_pi}}
\end{figure}

The importance of the double-scattering contribution of order $n^{4/3}$ to the in-medium pion mass is, of course, realized also in the chiral effective field theory approach \cite{ericson1966optical,waas1997deeply,goda2014pion}. In Fig.~\ref{fig:m_pi} we plot the FRG-ChNM model result for the pion mass as a function of density for symmetric nuclear matter at vanishing temperature. The non-trivial part of the corresponding curve starts at \mbox{$n = n_0$} because of the first-order liquid-gas transition. For comparison, the first-order (mean-field) approximation in the density expansion is shown, together with a recent in-medium chiral perturbation theory computation \cite{goda2014pion}. In agreement with ChEFT and phenomenology, we find an enhancement of the pion mass by about ten percent at nuclear saturation density.

As already noted in Ref.~\cite{drews2014functional}, we have not explicitly included an isospin chemical potential. Thus a potential source of isospin breaking is absent. The effect on the equation of state is expected to be negligible as was deduced from explicit calculations in chiral effective field theory \cite{kaiser2012isovector}. In contrast, this effect cannot be ignored when computing the in-medium pion mass for asymmetric nuclear matter. The masses of $\pi^+$, $\pi^-$ and $\pi^0$ split in such a medium \cite{kaiser2001systematic}. For example, the mass change for a $\pi^-$at leading order in the density (neglecting the small $b_0$ term) is now driven by the isospin-dependent parameter 
$b_1$: $\Delta m_\pi^{-}(n_n, n_p) \simeq -(2\pi/m_\pi)\,b_1\,(n_n-n_p)$, with $b_1 \simeq -0.1\, m_\pi^{-1}$. In neutron matter, the mass shift is repulsive for $\pi^-$ and attractive for $\pi^+$.

\section{Chiral symmetry restoration}\label{sec:chiral_restoration}
At low temperatures and small chemical potentials, chiral symmetry is spontaneously broken. At vanishing chemical potential, it is known from lattice calculations that chiral symmetry is restored in its Wigner--Weyl realization in a rapid crossover at temperatures above $T_c \simeq 155$ MeV \cite{bazavov2012chiral,borsanyi2010is}. It remains an open question whether this crossover turns into a first-order chiral phase transition for some positive chemical potential. If this were the case, there would exist a second order critical endpoint. Some model calculations based on effective quark degrees of freedom -- such as chiral quark-meson models or NJL type models -- predict a first-order transition at vanishing temperature for quark chemical potentials, $\mu_q$, around 300\,MeV (see e.g. \cite{roener2007polyakov,schaefer2007phase,fukushima2008phase,HRCW2009,herbst2013phase}). Translated into baryonic chemical potentials, $\mu_B\simeq3\mu_q$, chiral symmetry would be restored very close to the equilibrium point of normal nuclear matter, $\mu_B=923\text{\,MeV}$. Nuclear physics with its well-established empirical phenomenology teaches us that this can obviously not be the case. In this hadronic sector of the QCD phase diagram, nucleons and mesons are the predominantly active degrees of freedom, and chiral symmetry remains spontaneously broken in its Nambu--Goldstone realization.

We now explore chiral restoration in the ChNM model. The chiral order parameter in this model is the $\sigma$ field normalized to the pion decay constant $f_\pi$ in the vacuum. Its behavior in symmetric nuclear matter as a function of baryon density and temperature has been studied previously in refs.\,\cite{floerchinger2012chemical,drews2013thermodynamic}. In mean-field approximation \cite{floerchinger2012chemical}, chiral symmetry gets restored in a first-order transition at a density as low as $n \simeq 1.6\,n_0$. Once fluctuations are taken into account \cite{drews2013thermodynamic}, the chiral restoration transition is shifted to chemical potentials well beyond $\mu \sim 1$ GeV or densities exceeding  $3\,n_0$. Similar trends are found in ChEFT calculations of the in-medium chiral condensate,
\begin{align}
	\frac{\langle\bar\psi\psi\rangle(n,T)}{\langle\bar\psi\psi\rangle_0} = 1-\frac 1{f_\pi^2}\frac{\partial\mathcal F(n,T)}{\partial m_\pi^2}\,,
\end{align}
by computing the pion-mass dependence of the free-energy density,  $\mathcal F(n,T)$ \cite{fiorilla2012chiral, KHW2008}. 

\begin{figure}
	\centering
	\begin{overpic}[width=0.45\textwidth]{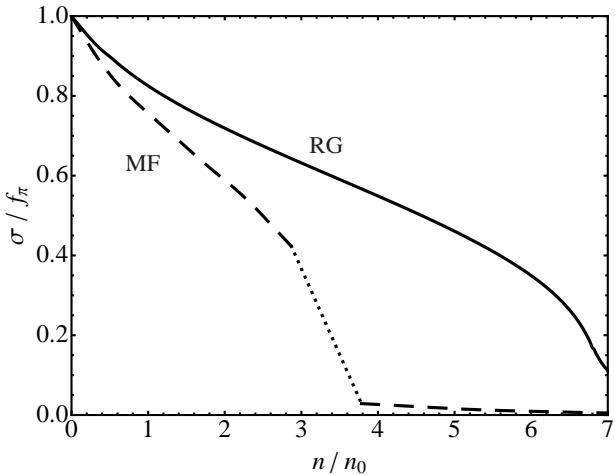}
		\put(20,50){MF}
		\put(50,53){RG}
	\end{overpic}
	\vspace{-0.2cm}
	\caption{Density dependence of the chiral order parameter at vanishing temperature for pure neutron matter. Solid line (RG): FRG-ChNM model calculation. Dashed line (MF): mean-field result.}
	\label{fig:chiral_restoration_NM}
\end{figure}

The chiral condensate in pure neutron matter has been calculated previously within (perturbative) chiral effective field theory \cite{kaiser2009chiral,krueger2013chiral}. Chiral nuclear forces treated up to four-body interactions at N${}^3$LO \cite{krueger2013chiral} were shown to work moderately against the leading linear fall of the condensate with increasing density around $n \simeq n_0$, with a rate of stabilization less pronounced than in symmetric nuclear matter. It is instructive at this point to investigate the density dependence of the chiral order parameter $\sigma$ in neutron matter within the FRG-ChNM model. The non-perturbative FRG approach permits extrapolating to higher densities.  Results are presented in Fig.\,\ref{fig:chiral_restoration_NM}. In mean-field approximation the order parameter $\sigma/f_\pi$ shows a first-order chiral phase transition at a density of about $3\,n_0$. However, the situation changes qualitatively when fluctuations are included using the FRG framework. The chiral order parameter now turns into a continuous function of density, with no indication of a phase transition. Even at five to six times nuclear saturation density the order parameter still remains at about forty percent of its vacuum value. Only at densities as large as $n \sim 7n_0$ does the expectation value of $\sigma$ show a more rapid tendency towards a crossover to restoration of chiral symmetry in its Wigner--Weyl mode. We thus observe a huge influence of higher order fluctuations involving Pauli blocking effects in multiple pion-exchange processes and multi-nucleon correlations at high densities. With neutron matter remaining in a phase with spontaneously broken chiral symmetry even up to very high densities, this encourages a further-reaching application and test of the FRG-ChNM model by constructing an equation of state for the interior of neutron stars. 

\section{Neutron stars}\label{sec:neutronstars}

\begin{figure}
	\centering
	\begin{overpic}[width=0.45\textwidth]{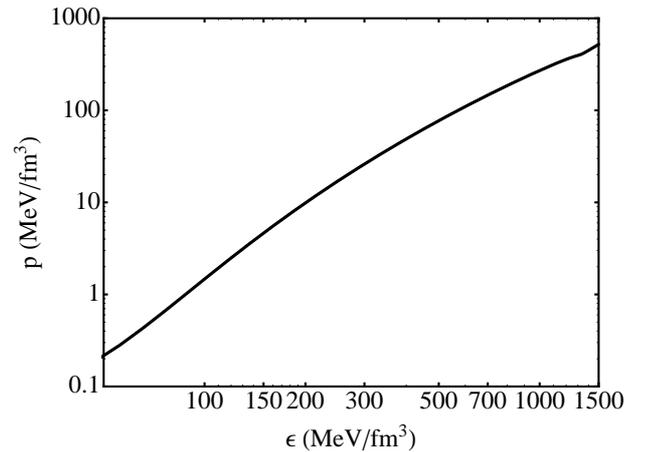}
	\end{overpic}
	\vspace{-0.2cm}
	\caption{The equation of state (pressure versus energy density) of neutron matter calculated within the FRG-ChNM model, taking beta equilibrium into account and using \mbox{$E_{\text{sym}}=32\text{ MeV}$}.} 
\label{fig:EoS}
\end{figure}

Neutron stars are electrically neutral objects. To provide negative charges, the nucleon-meson model has to be extended by inclusion of electrons and muons. The condition of charge neutrality is
\begin{align}
	n_p=n_e+n_\mu\,.
\end{align}
Moreover, matter in the interior of a neutron star is subject to chemical beta equilibrium. Neutron, proton, and electron chemical potentials are related by
\begin{align}
	\mu_p+\mu_e=\mu_n\,.
\end{align}
Electrons and muons are assumed to contribute as free Fermi gases to the energy density and pressure of the model. Charge neutrality and beta equilibrium leave only one single chemical potential as a free parameter.  With these conditions an equation of state applicable to the interior of a neutron star is readily calculated within the FRG-improved ChNM model. In Fig.~\ref{fig:EoS}, the pressure is shown as a function of the energy density.

Given an equation of state $p(\epsilon)$, the mass and radii of neutron stars can be computed from the Tolman--Oppenheimer--Volkoff (TOV) equations \cite{tolman1934effect,tolman1939static,oppenheimer1939massive}
\begin{align}
	\begin{aligned}
\frac{dp(r)}{dr}&=-\frac G{r^2}\left[\epsilon(r)+p(r)\right]{M(r)+4\pi r^3p(r)\over 1-2GM(r)/r}\,, \\
		&~~~~~~~\frac{dM(r)}{dr}=4\pi r^2\epsilon(r)\,.
	\end{aligned}
\end{align}
Here $G$ is the gravitational constant and $r$ is a radial parameter. The boundary conditions at $r = 0$ are $M(0)=0$ and $\epsilon(0)=\epsilon_c$, where the central energy density $\epsilon_c$ is varied to generate a mass-radius curve.

The outer crust of the neutron star consists of an iron lattice and hence the energy density is that of iron, \mbox{$\epsilon_{\stext{Fe}}=4.4\cdot 10^{-12}\text{ MeV/fm}^3$}. The neutron star radius is then implicitly defined by the relation $\epsilon(R)=\epsilon_{\stext{Fe}}$. Finally, the mass is obtained from the TOV equations as
\begin{align}
	M=M(R)=4\pi\int_0^Rdr\;r^2\epsilon(r)\,.
\end{align}
Moving inwards from the crust, the nuclei become more neutron rich as the density increases and electrons are captured \cite{chamel2008physics}. The inner crust contains (possibly superfluid) neutrons. The crust is frequently parametrized by the Skyrme-Lyon (SLy) equation of state \cite{baym1971ground,douchin2001unified}. This SLy EoS is matched to the FRG-ChNM model at the point where the energy-density curves intersect, which happens at a density $n\simeq0.3\,n_0$. From there on to higher densities the FRG-ChNM equation of state is taken as a model for the neutron star core. We do not consider a possible transition to quark matter, nor do we include other exotic types of matter such as kaon condensates. Hyperons are also not included as they would generally soften the equation of state unless strong additional repulsion is introduced for compensation \cite{weissenborn2012hyperons,lonardoni2014hyperon}.

\begin{figure}
	\centering
	\begin{overpic}[width=0.45\textwidth]{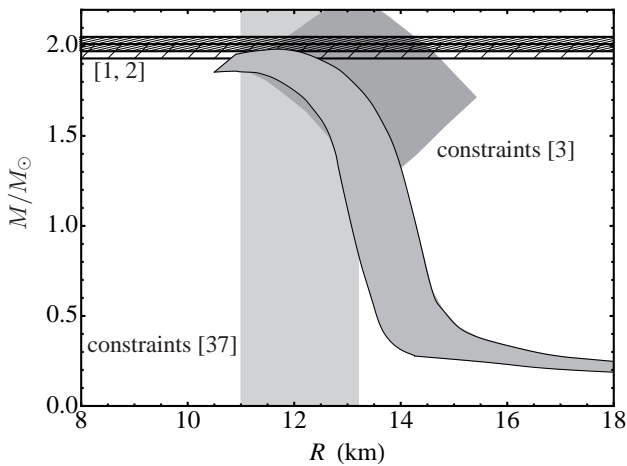}
		\put(13,65){\cite{demorest2010two-solar-mass,antoniadis2013massive}}
		\put(69,52){constraints \cite{truemper2011observations}}
		\put(12,20){constraints \cite{lattimer2014constraints}}
		\put(2,40){\begin{rotate}{90} $M/M_{\odot}$ \end{rotate}}
	\end{overpic}
	\vspace{-0.2cm}
	\caption{Mass radius relation for neutron star matter. The mass-radius constraints from reference \cite{truemper2011observations}, the radius constraint~\cite{lattimer2014constraints} and the two-solar-mass neutron stars \cite{demorest2010two-solar-mass,antoniadis2013massive} are shown for comparison.\label{fig:m_r}}
\end{figure}
\begin{figure}
	\centering
	\begin{overpic}[width=0.45\textwidth]{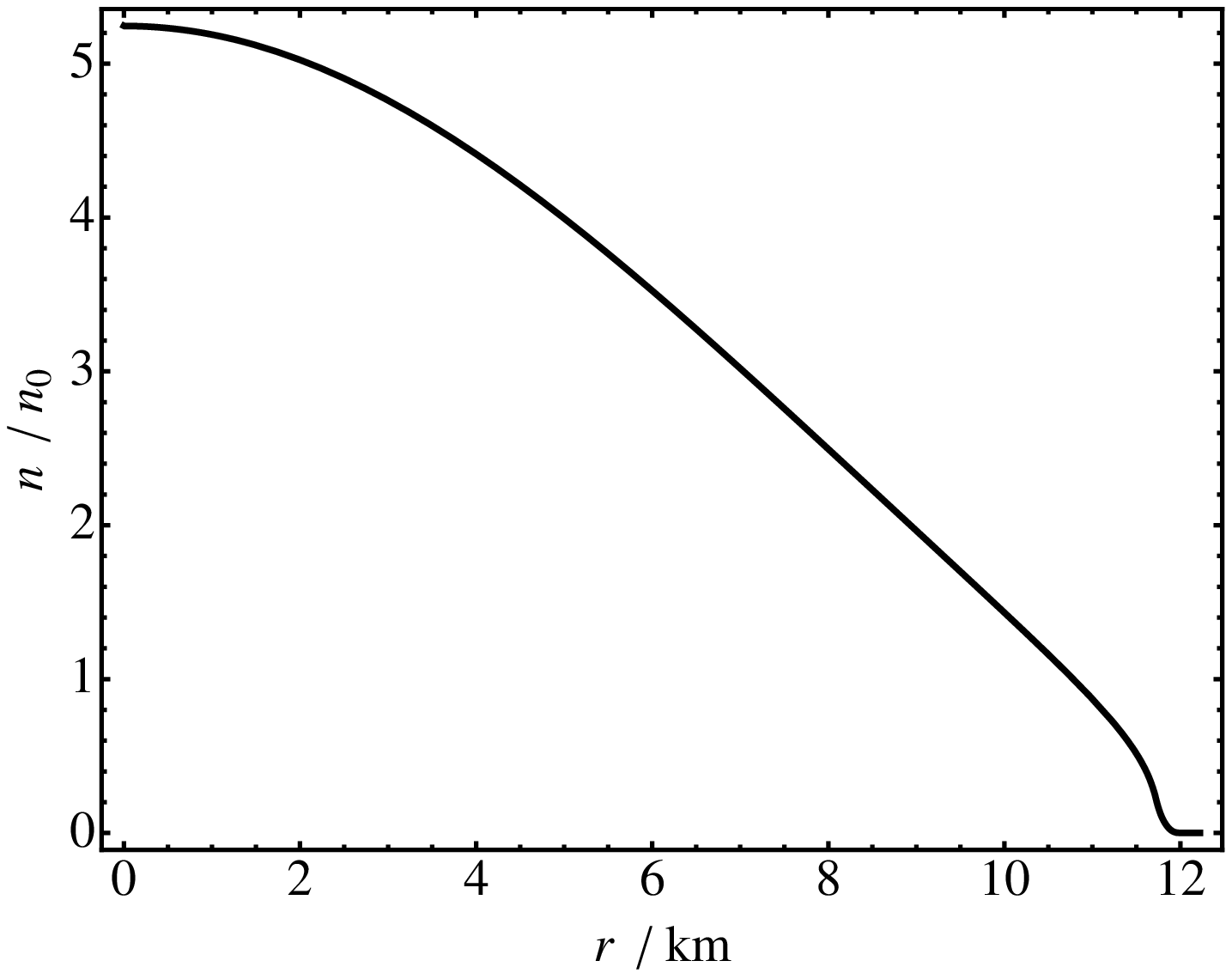}
	\end{overpic}
	\vspace{-0.2cm}
	\caption{Density profile for a neutron star with mass \mbox{$M=1.97\;M_\odot$} and \mbox{$R=12.2\text{ km}$} for \mbox{$G_\rho=1.46\text{ fm}^{-2}$}. \label{fig:n_r}}
\end{figure}

In Fig.~\ref{fig:m_r} the mass-radius relation of the FRG-ChNM model obtained from the TOV equations is shown. The gray band of mass-radius trajectories results from using symmetry energies in an empirically acceptable range \mbox{$29\text{ MeV}\le E_{\stext{sym}}\le37\text{ MeV}$} (or, correspondingly, a range of isovector-vector couplings \mbox{$0.91\text{fm}^2\le G_\rho\le1.46\text{ fm}^2$}). For not too small symmetry energies the equation of state is found to be in agreement with the observed two-solar-mass neutron stars.

Unlike the precise $2\,M_{\odot}$ mass determinations, the radius constraints for neutron stars are far less accurate. They are subject to model dependent assumptions. Nevertheless, limits on minimal and maximal radii can be inferred from different sources, such as X-ray burst oscillations, thermal emission, and stars with largest spin frequency. The result of such a detailed analysis \cite{truemper2011observations} is a rhomboidal region (depicted in gray in Fig.~\ref{fig:m_r}), which a realistic equation of state must intersect. Our equation of state is in agreement with all these constraints. For comparison the acceptable radius interval according to Ref.~\cite{lattimer2014constraints} is also shown.

Figure \ref{fig:n_r} shows a typical calculated density profile of a neutron star with mass $M=1.97\;M_\odot$ using \mbox{$G_\rho=1.46\text{ fm}^{-2}$} (i.e. a symmetry energy of 37 MeV). It is noteworthy that even in the center of the neutron star the density does not exceed about five times nuclear saturation density. The required stiffness of the EoS does not permit ultrahigh densities in the inner core of the star. These findings are in agreement with a corresponding ChEFT computation \cite{hell2014dense}.

The question might nonetheless be raised whether approaches such as the ChNM model, based entirely on spontaneously broken chiral symmetry with pions and nucleons as degrees of freedom, are still applicable at densities as high as $5\,n_0$. Clearly, the mean-field version of the ChNM model with its first-order chiral phase transition at about $3\,n_0$ would not qualify for such extrapolations. The FRG-improved version of this model on the other hand, with proper non-perturbative treatment of fluctuations and many-body correlations, is prepared to deal with dense baryonic matter. Even if the nucleon mass at $n \sim 5\,n_0$ is reduced to less than half of its vacuum value, chiral symmetry is still realized in the spontaneously broken Nambu--Goldstone phase at such densities, with no trace of a nearby chiral phase transition.

\section{Summary and conclusions}
A chiral nucleon-meson model, previously designed to describe isospin-symmetric nuclear matter, has been extended to asymmetric nuclear matter and neutron matter. Fluctuations beyond the mean-field approximation are treated using functional renormalization group methods. All parameters have been fixed to properties of symmetric nuclear matter together with the empirical symmetry energy. 

The behavior of the nuclear liquid-gas phase transition has been studied in detail. The critical endpoint is in excellent agreement with empirical data. The equations of state of symmetric, asymmetric and pure neutron matter have been computed. With only one additional free parameter (fitted to the symmetry energy), the equation of state of neutron matter is in remarkable agreement with advanced many-body computations for densities up to several times nuclear saturation density. Likewise, the equation of state of symmetric nuclear matter is well reproduced in comparison with realistic many-body calculations. 

The in-medium behavior of the pion mass has been analyzed as a test example for the treatment of fluctuations, in this case involving dominant double-scattering mechanisms of the pion in nuclear matter. It is found that the pion mass is enhanced by about ten percent in isospin-symmetric matter at nuclear saturation density, consistent with phenomenology and in-medium chiral perturbation theory.

Moreover, it has been demonstrated that fluctuations beyond mean-field approximation play an extremely important role in stabilizing the chiral order parameter as a function of density, both in symmetric nuclear matter and neutron matter. We find that chiral symmetry remains spontaneously broken in neutron matter up to at least six times nuclear saturation density. The resulting equation of state in the FRG-ChNM model turns out to be sufficiently stiff in order to satisfy the updated constraints imposed by neutron star observations. 

\section*{Acknowledgments}
This work is supported in part by BMBF and by the DFG Cluster of Excellence ``Origin and Structure of the Universe.'' One of us (W.W.) gratefully acknowledges inspiring discussions with Christof Wetterich.

\bibliography{biblio}{}

\end{document}